\RequirePackage{fix-cm}
\documentclass[smallextended]{svjour3}       
\smartqed  
\usepackage{graphicx}
\begin{document}

\title{Intrinsic vanishing of energy and momenta in a universe
}


\author{Ramon Lapiedra \and Juan Antonio Morales--Lladosa}


%
\institute{Ramon Lapiedra \and Juan Antonio Morales--Lladosa \at Departament
d'Astronomia i Astrof\'{\i}sica,
\\Universitat
de Val\`encia, E-46100 Burjassot, Val\`encia, Spain.\\
Tel.: +34-96-3543066\\
Fax: +34-96-3543084\\
\email{ramon.lapiedra@uv.es}\\
\email{antonio.morales@uv.es}}

\date{Received: date / Accepted: date}

\maketitle

\begin{abstract}
We present a new approach to the question of properly defining
energy and momenta for non asymptotically Minkowskian spaces in
General Relativity, in the case where these energy and momenta are
conserved. In order to do this, we first prove that there always
exist some special Gauss coordinates for which the conserved
linear and angular 3-momenta intrinsically vanish.  
This allows us to consider the case of {\it creatable}
universes (the universes whose proper 4-momenta vanish) in a
consistent way, which is the main interest of the paper. When
applied to the Friedmann-Lema{\^{\i}}tre-Robertson-Walker case,
perturbed or not, our formalism leads to previous results,
according to most literature on the subject. Some future work
that should be done is mentioned.
\end{abstract}


\section{Introduction}
\label{intro}

\subsection{General considerations}

In General Relativity, the problem of associating linear and angular 
4-momenta to a finite space-time region and
the related idea to define (for a general space-time) such global
quantities have been approached from different, but not necessarily contradictory,
perspectives. See \cite{Szabados-review} for an extensive and critical
review on the current status of the problem.

Diverse mathematical objects (energy-momentum pseudotensors and
superpotentials, flat or curved  background
metrics, Killing vectors or other fields generating generalized
symmetries, etc.) and several geometrical techniques (3+1 or 2+2
space-time splittings, initial data constraints and boundary
conditions, etc.) seem appropriate to deal with this issue. See, for example, 
\cite{Katz-Bi-LyBell-97,Nester-2004,Brown-York-93,Hawking-Horowitz-96,Silva} 
for some detailed explanations and general comments on
these subjects.

However, nowadays, no consensus on a preferred
approach nor any complete or definitive answer to the problem of
how to associate linear and angular 4-momenta to a general
space-time seem to have been reached by the relativistic
community.  Of course, the existing points of view don't exclude each other 
and seem to point towards the correct understanding of the problem, while the
possibility of new approaches remains still open.

Nevertheless, the reader should be warned about the presence of a lot of criticisms in the current literature to the pseudotensor approach to define the 4-momenta of a physical space-time, which is the approach adopted in the present paper. These criticisms stress that the approach is by no means a covariant one (see for example \cite{FatibeneFF,BibbonaFF}), or argue against any definition of energy in General Relativity referring to 3-surface integrals, instead of being quasi-local (referring to 2-surface integrals) from the very beginning  \cite{Hayward}, or even accept the approach for asymptotically flat space-times but express some doubts for the non asymptotically flat ones \cite{Cooperstook}.

Although the covariant approach to the definition of quasi-local conserved quantities in General Relativity followed in \cite{FatibeneFF,BibbonaFF,Hayward} and the results obtained  seem very interesting, we cannot fully share those criticisms.

Before giving our personal opinion about it, let us begin remarking that in \cite{Chang-Nester-Chen-1999} a particular ``covariant Hamiltonian approach to quasi-local energy'' is presented. In this approach, each pseudotensor corresponds to a Hamiltonian boundary term, which brings the authors to the conclusion that ``Hamiltonian approach to quasi-local energy-momentum rehabilitates the pseudotensors''. Quoting this conclusion, Vargas \cite{Vargas-2004} used again the pseudotensorial method to calculate the energy of the universe in teleparallel gravity. The conclusion was quoted in  \cite{Chen-Liu-Nester-2007} too.

Regarding the objection prescribing a quasi-local definition of energy in General Relativity \cite{Hayward}, we recall that our pseudotensorial method yields 4-momenta that are quasi-local quantities, in the sense that they can be expressed subsequently as 2-surface integrals, even if  their original definitions were through 3-surface integrals.

As far as the remarked \cite{FatibeneFF,BibbonaFF} non covariance of the pseudotensorial method is concerned, we recall that there is nothing invalidating in the fact that the energy of a physical system can depend on the reference frame used and that, at the same time, we look for some natural special frame in order to define some ``proper'' energy and momenta: at least, nothing, apparently, that from the very beginning prevents us from approaching this problem. The same frame dependence is present in classical mechanics, as it is mentioned in \cite {CariniFF} in relation to the general problem of defining energy in a covariant way. But, for a particle, for example, we can select a ``natural'' special frame (the one where the 3-momentum vanishes) to define the ``proper'' energy of the particle. In a similar way, but pointing to General Relativity, we must select a ``natural'' congruence of observers and a ``natural'' coordinate system related to it in order to get rid of the spurious energy and momenta associated to the fictitious gravitational field related to ``bad'' observers (for example, observers which do not fall freely) and ``bad'' related coordinates, so that we can reach some kind of  space-time ``proper'' 4-momenta (see section \ref{sec-2}, in the paragraph beginning just after Eqs. (\ref{energy})-(\ref{angular time momentum})). This is in fact what is performed in the non problematic case of asymptotically flat space-times, where the ``proper'' coordinates are those which are asymptotically Lorentzian, such that the corresponding energy is the ``proper'' energy. We can think about the pseudotensorial method developed in the present paper as a proper generalization of this procedure to the non asymptotically flat case, which would be a certain response to the aforementioned doubts in \cite{Cooperstook}. In fact, in \cite{Szabados-review} Szabados summarizes the question by simply saying: ``to use the pseudotensors, a ``natural'' choice for a ``preferred'' coordinate system would be needed''. This is just what we have done in the present paper.

\subsection{Summary of some previous work}

In a previous paper \cite{Ferrando}, we addressed the
question of properly defining the linear and the angular 4-momenta
of a significant family of non asymptotically flat space-times. As
it is well known, and as we have just commented, see for example \cite{Weinberg} or
\cite{Murchadha}, this proper definition can be accomplished
without difficulty in the opposite case of asymptotically flat
space-times, but not in the general case (for a concise and readable account, 
see also \cite{Alcubierre}). The reason for this
difficulty in the general case stays in the dramatic dependence of
these momenta on the coordinate system used. This fact is very
well known but very few times has properly been taken into account
in the literature of the field, where some authors use a given
coordinate system to calculate some of the momenta, without any
comments on the rightness of the coordinate selection that has
been done. For related questions on this subject see, for
instance, \cite{Katz-Bi-LyBell-97,Banerjee,Xulu,Nester-So-Va-08,Pitts-10} and references therein.

The family of space-times that we are going to consider in the
present paper is the family of all non asymptotically flat
space-times where these well defined momenta are conserved in
time. We call these particular space-times \emph{universes}, since
it is to be expected that any space-time which could represent the
actual universe should have conserved momenta, provided that these
momenta be properly defined, which is the goal achieved in the
present paper.

Then like in \cite{Ferrando}, we call \emph{creatable universes} the universes
which have vanishing $4$-momenta, since again this is what could
be expected to happen if the considered universe raised from a
quantum fluctuation of the vacuum \cite{Albrow,Tryon}. In fact,
the question of the \emph{creatable universes} is our main
motivation to consider the subject of properly defining the
momenta of non asymptotically flat space-times. Demanding the
vanishing of the momenta can be a way of saying something relevant
about how our actual Universe looks like either now or in the
preinflationary phase. Thus, for example, in \cite{Lapiedra-Saez},
perturbed flat Friedmann-Lema{\^{\i}}tre-Robertson-Walker (FLRW)
universes according to standard inflation, and also the perturbed open universes,  
were found to be non creatable. Therefore, among the inflationary perturbed FLRW
universes, only the closed ones would be left as good candidates
to represent the actual Universe.

In the present paper we present a new approach to the subject of properly
defining the two $4$-momenta  of a {\em universe}, as compared
with the one presented in the above reference \cite{Ferrando}
whose results we summarize here:  

In \cite{Ferrando} we considered a given space-time, not necessarily asymptotically flat, with its general expressions for the linear 4-momentum and the angular 4-momentum obtained from the Weinberg complex. Then, we assumed that the ``intrinsic'' values of these 4-momenta are conserved, that is, the values corresponding to some ``proper'' coordinate system that has to be consequently determined. As mentioned above, space-times endowed with such conserved 4-momenta are called {\it universes} in the referred paper. We argued why these coordinates, $\{t, x^i\}$, are, to begin with, Gauss coordinates referred to some space-like 3-surface, $\Sigma_3$, whose equation then becomes $t=t_0$. We proved that the corresponding 3-space metric, $dl^2_0$, that is, $dl^2 = g_{ij} dx^i dx^j$ for $t=t_0$, is asymptotically conformally flat over the 2-surface boundary, $\Sigma_2$, of $\Sigma_3$, and we used 3-space coordinates $x^i$ adapted to this circumstance, i.e., $dl^2|_{\Sigma_2}= f \delta_{ij} dx^i dx^j$, with $f$ some function defined on $\Sigma_2$. Finally, looking for universes with vanishing 4-momenta, we assumed that the metric components $g_{ij}$ go to zero fast enough when we approach $\Sigma_2$. In this way, we were able to define a family of universes whose 4-momenta vanish irrespective of the selected $\Sigma_3$ and of the conformal coordinates used in the corresponding boundary  $\Sigma_2$. The family covers in particular the FLRW universes, for which we obtain the 4-momenta values previously obtained by some authors but not by all of them (see section \ref{sec-6} for some comments about these agreements and disagreements). 

The new approach is by no means a minor variation of the ancient one, as we explain in the next subsection:

\subsection{Outline of the paper}

In the present paper, given a {\em universe}, when trying to select the appropriate
coordinate systems in order to properly define its two
$4$-momenta, $P^\alpha$ and $J^{\alpha \beta}$, we impose
alternatively to \cite{Ferrando} that both $3$-momenta, $P^i$ and
$J^{ij}$, vanish, the last one irrespective of the origin of
momentum. However, according to what we have just explained about \cite{Ferrando}, we rest on Gauss
coordinates based on some space-like $3$-surface, $\Sigma_3$, such
that the corresponding $3$-space metric can be written in a
conformally flat way on the boundary of $\Sigma_3$. Such Gauss
coordinate systems, where both $3$-momenta vanish (the last one
irrespective of the origin), which at the same time are coordinates satisfying the above conformally flat property, will
be called here {\em intrinsic} coordinate systems. Obviously, we first 
prove here that these {\it intrinsic} coordinate systems always
exist for any {\em universe}, which is a capital new result.

However, in \cite{Ferrando}, in order to have vanishing $4$-momenta, we had to assume that
the metric and its first derivatives went fast enough to zero when
we approach the boundary of $\Sigma_3$. In the present paper we do
not need to make such an {\it ad hoc} assumption, and so our present approach, 
as compared with the one in \cite{Ferrando}, stresses the {\it intrinsic} character, 
and so the physical meaning, of the given definition of the universe $4$-momenta.

The paper is organized as follows:  In Sect.  \ref{sec-2}, given
a space-like $3$-surface, $\Sigma_3$, we give the corresponding
family of coordinate systems where to choose the right coordinate
systems to properly define the linear and the angular $4$-momenta
associated to this $\Sigma_3$. In section \ref{sec-3}, we consider
all $3$-surfaces $\Sigma_3$ showing the same boundary $\Sigma_2$.
Then, by defining what we have called intrinsic coordinates, we
select the $3$-surfaces $\Sigma_3$ for which the linear and the
angular $3$-momenta vanish, after proving that this result is
valid for some $\Sigma_3$. In Sect. \ref{sec-4}, we define
the notion of {\it creatable} universe and we discuss briefly its
goodness. In Sect.  \ref{sec-5}, we invoke some previous results
to check the {\it creatibility} of the perturbed FLRW models in the new
scheme, reproducing the known conclusions  also obtained in \cite{Lapiedra-Saez} 
on these models. Had we not been able to confirm these results  
in the present approach, we should consider them as actually non valid, since we find now that 
the vanishing of $P^i$ and $J^{ij}$ in the coordinates used is mandatory (although not sufficient) to confer physical meaning to the $4$-momenta definition used. Finally, in Sect. \ref{sec-6}, we comment on the, sometimes, different values of the 4-momenta for FLRW universes, found by different authors, including our work, and we point out which is, in our opinion, the main interest of the paper, and in relation to this we refer to some
future work.

We still add three appendices where some calculations are given in
detail.

A short report containing some results, without proof,  of this
work was presented at the Spanish Relativity Meeting
ERE-2009 \cite{ere09}.

\section{The energy and momenta of a \emph{universe}, associated to a given space-like 3-surface}
\label{sec-2}
In order to define the linear and angular 4-momenta of a
\emph{universe} we will use the Weinberg complex \cite{Weinberg}.

It remains to be checked wether the final results obtained in the
present paper keep still valid for other complexes that, like the
Weinberg one, are symmetric in their two indices, which allows us to build 
the corresponding angular 4-momentum. This criterion leads to discard other 
pseudotensors as the ones by Einstein, Bergmann or M{\o}ller, but not 
by Papapetrou or Landau-Lifshitz. Any case, the Weinberg complex
is a very natural one, as it is very convincintly argued in 
\cite{Weinberg}, letting aside the interesting well known fact
that Weinberg complex gives also the correct 4-momenta in a
Schwarzschild metric. As far as the Golberg pseudotensor is concerned, 
it is a very general one to which the Weinberg one belongs as a particular case. 
About these different pseudotensor see, for example \cite{Chang-Nester-Chen-1999}.

Before going to the notion of the 4-momenta of a general space-time, some previous definitions and considerations.

Metric signature: we use signature $+ 2$, that is, $d \tau^2 \equiv  - ds^2 = - g_{\alpha\beta} dx^\alpha dx^\beta$ is the square of the corresponding elementary proper time when $ds^2 < 0$. Thus, Greek indices take values from 0 to 3, and Latin indices from 1 to 3. 

Gauss coordinates: we can define them as coordinates in which $ds^2 = - d t^2 + d l^2$ with 
$dl^2 \equiv g_{ij} dx^i dx^j$ being positive defined. Although not globally, we can build such a coordinate system by referring the space-time metric to a congruence of observers that fall freely, by endowing each observer with a canonical (physical) clock, and finally by {\it synchronizing} (see next the notion of synchronization) all these different clocks \cite{Landau}. In these coordinates, the 3-surface $t=t_0$, with $t_0$ any constant time, is a space-like 3-surface,  $\Sigma_3$, orthogonal to the congruence, in whose neighborhood the Gauss coordinate system is defined.

Clock synchronization: given such a congruence of observers, each one endowed with his canonical clock, we can synchronize all them with the same method used in special relativity (using come and back light beams: see again \cite{Landau}). Then, one finds that all events belonging to $\Sigma_3$, that is, all the events $t=t_0$, are simultaneous according to this definition. Thus, $t$ is a physical and universal time (like time in the Minkowski space is, for example).

Then, to properly define the notion of 4-momenta of a \emph{universe},
associated to some space-like 3-surface, $\Sigma_3$, we will take
Gauss coordinates associated to this 3-surface, $\Sigma_3$, in the
neighborhood of it (we explain next why we make this choice).
In the Weinberg approach \cite{Weinberg}, the linear and angular momenta of the gravitational field are incidentally defined by integrating on $\Sigma_3$. The main Weinberg pursuit is to obtain an integral balance relation for each momentum component such that, as it is standard, the time derivative of a 3-volume $\Sigma_3$ integral for this component density equates (using Gauss theorem) the minus correspondent flux through the 2-surface boundary $\Sigma_2$ of the 3-volume $\Sigma_3$. These integrated balance equations come straightly from the vanishing of the ordinary (not covariant) divergence of the pseudotensor plus the energy-momentum tensor and, by construction, the 3-volume integrals incorporate a non geometric volume element. That is, this 3-volume element is just $dx^1 dx^2 dx^3$ independently of the meaning of this coordinates (see the details in \cite{Weinberg}). As a result, these volume integrals can be expected to have a physical meaning only for some kinds of physical coordinates. On the other hand, these 3-volume integrals, using again Gauss theorem, can be written as 2-surface integrals over the 3-volume boundary, $\Sigma_2$. Then, according to \cite{Weinberg}, we have for the corresponding
energy, $P^0$, linear 3-momentum, $P^i$, angular 3-momentum,
$J^{ij}$, and components $J^{0i}$ of the angular 4-momentum, of
the \emph{universe}:
\begin{eqnarray}
P^0 & = & \kappa \int(\partial_j g_{ij} - \partial_i g)
d \Sigma_{2i},  \label{energy}\\[3mm]
P^i & = & \kappa \int(\partial_0{g} \delta_{ij} -
\partial_0{g}_{ij}) d \Sigma_{2j},
\label{linear momentum}\\[3mm]
J^{jk} & = & \kappa \int(x_k \partial_0 {g}_{ij} - x_j
\partial_0 {g}_{ki}) d \Sigma_{2i},\label{angular momentum}
\\[3mm]
J^{0i} & = & P^i t - \kappa \int[(\partial_k g_{kj} -
\partial_j g)x_i + g \delta_{ij} - g_{ij}] d \Sigma_{2j}, \quad
\label{angular time momentum}
\end{eqnarray}
where we have used the following notation: $\kappa^{-1} \equiv 16
\pi G$, $G$ is the Newton constant and we have taken $c=1$ for the speed of light, $g \equiv \delta^{ij}g_{ij}$,
$\partial_0$ is the partial derivative with respect to $x^0 \equiv
t$, and $d \Sigma_{2i}$ is the surface element of
$\Sigma_{2}$, the boundary of $\Sigma_3$. Further, indices $i, j,
... $ are raised or lowered with the Kronecker $\delta$ and
angular momentum has been
taken with respect to the origin of coordinates.%
\footnote{In the Weinberg book \cite{Weinberg}, the case of an
asymptotically flat space-time is the only considered.
Nevertheless, it is straightforward to see that the displayed
treatment also covers the case of non asymptotically flat ones,
provided that $P^\alpha$ and $J^{\alpha\beta}$ as defined in
(\ref{energy})-(\ref{angular time momentum}) exist, i.e., provided
that the corresponding integrals converge. \label{fn1}}

Why Gauss coordinates? We expect any well behaved universe, $V_4$,
to have well defined energy and momenta, i. e., $P^{\alpha}$ and
$J^{\alpha \beta}$, $\alpha, \beta, ... =0,1,2,3$, such that they
are finite and conserved in time (a \emph{universe} in our
notation). So, for this conservation to make physical sense, we
need to use a \emph{physical} and \emph{universal} time, according to the definition introduced at the beginning of the present Section. Then, still in accordance with these definitions,  we are conveyed to use a Gauss coordinate system with its universal time to properly define the \emph{universe} 4-momenta. Moreover, using this Gauss time, any component of the $4$-momenta appears as the addition (the $2$-surface integral) of the corresponding {\it simultaneous} densities, as it must be from a physical point of view. 

Then, as defined above, we will have for the line element of $V_4$:
\begin{equation}
d s^2 = -dt^2 + dl^2, \quad dl^2 = g_{ij} dx^i dx^j,
\label{4-metric}
\end{equation}
and we can write $t=t_0 = constant$ for the equation of
$\Sigma_3$.

The area of the 2-surface boundary $\Sigma_2$ could be zero,
finite or infinite. Let us precise that in the first case, when
the area is zero, the 4-momenta do not necessarily vanish, unless
the metric and its first derivatives remain conveniently bounded
when we approach $\Sigma_2$.

Obviously, we have as many local families of Gauss coordinates as
space-like 3-surfaces, $\Sigma_3$, we have in $V_4$. Then,
$P^{\alpha}$ and $J^{\alpha \beta}$ will depend on $\Sigma_3$,
which is not a drawback in itself (the energy of a physical system
in the Minkowski space-time also depends on the $\Sigma_3$ chosen,
i.e., on the Lorentzian coordinates chosen). But the problem is
that, given a space-like 3-surface, $\Sigma_3$, we can still have
many different 4-momenta, according to the particular Gauss
coordinate we choose, associated to the same $\Sigma_3$.

Let us begin suppressing a part of the arbitrariness left in the choice of
Gauss coordinates. In order to do this, we will choose Gauss coordinates
such that the equation of $\Sigma_2$ becomes $x^3=0$, and $dl^2$ on
$\Sigma_2$ reads
\begin{equation}\label{3-metric}
dl^2 (t=t_0, x^3=0) \equiv dl^2|_{\Sigma_2} = f(x^a) \delta_{ij}
dx^i dx^j,
\end{equation}
with $f$ some given function,%
\footnote{Expression (\ref{3-metric}) will not always be valid globally. In this case we will have to cover $\Sigma_2$ with different charts, performing each $\Sigma_2$--integration over each chart, and summing up the different non overlapping chart contributions. \label{fn2}}
$a,b, ... = 1,2$, and furthermore
\begin{equation}\label{g3a}
g_{3a}(t=t_0) = 0.
\end{equation}
That can always be done (see \cite{Ferrando} and the last paragraph of the present section). 
Therefore, the induced 3-volume element $dx^1 dx^2 dx^3$ used in our 3-volume integrals to define energy and momenta (see the just previous paragraph to Eqs.  (\ref{energy})-(\ref{angular time momentum})) becomes physically sound.

Furthermore, since $t=t_0$, $x^3 = 0$, is now the equation of the
2-surface $\Sigma_2$,  the expressions (\ref{energy})-(\ref{angular
time momentum}) for $P^\alpha$ and $J^{\alpha\beta}$ simplify to:
\begin{eqnarray}
P^0 & = & - \kappa \int \partial_3 g_{aa} \ d x^1 d x^2 \, , \label{P0}\\
P^a & = & - \kappa \int \partial_0 g_{3a} \ d x^1 d x^2 \, , \label{Pa}\\
P^3 & = & \kappa \int \partial_0 g_{aa} \ d x^1 d x^2 \, , \label{P3} \\
J^{ij} & = & \kappa \int(x_j \partial_0
g_{3i} - x_i \partial_0 g_{3j}) \ dx^1 d x^2 \, , \label{Jij}\\
J^{0a} & = &  P^a t_0 + \kappa \int x^a  \partial_3 g_{bb} \  d x^1 d x^2 \, , \label{J0a}\\
J^{03} & = & P^3 t_0 - \kappa \int g_{aa} \ d x^1 d x^2 \,
\label{J03}
\end{eqnarray}
where $g_{aa} = g_{11} + g_{22}$. Notice that, since the 3-volume element was  $dx^1 dx^2 dx^3$, corresponding to the 3-metric 
$\delta_{ij} dx^i dx^j$, and since $\Sigma_2$ is $x^3=0$, the induced 2-metric on $\Sigma_2$ is $\delta_{ab} dx^a dx^b$, whose metric determinant value is $1$, and so the 2-surface element $d \Sigma_{2i}$ has become $dx^1 dx^2$.

Let us point that $\Sigma_2$ could also be made of different sheets. Thus, in the above
Gaussian coordinates,  these sheets could be the six faces of a
cube that increases without limit. Its corresponding six equations
would be $\forall i, x^i = \pm L$,  for $L \to \infty$. These
equations could be written $x'^i = 0$, by defining the new
coordinates $x'^i = x^i \mp L$ and putting $x'^i = \pm
|\epsilon|$, with $L \to \infty$ and $\epsilon \to 0$, that is, we
will first calculate the integrals (\ref{energy})-(\ref{angular
time momentum}) for finite values of $L$ and $|\epsilon|$, and
then we will take the above limits. The new coordinates $x'^i$ are Gauss coordinates with 
$dl'^2$ conformally flat on $\Sigma_2$ as it must be. All this means, in particular, that the right hand side of, for example, (\ref{P0}) would
actually be the sum of six similar integrals, one for each cube
face. Nevertheless, in case we would have taken $x^i = \pm L$, $L \to \infty$,
as the equation of $\Sigma_2$, it can be easily seen that 
essentially nothing would change in the present paper.

\section{Proving that, for any \emph{universe}, intrinsic coordinates always exist}
\label{sec-3}

We start with a Gauss coordinate frame, $\{x^\alpha\}$, such that
(\ref{3-metric}) and (\ref{g3a}) are satisfied. Let us prove that,
from this coordinate frame, we always can move to an {\em
intrinsic} coordinate frame as defined in the Introduction. Let it
be a coordinate transformation $x^\alpha \to x'^\alpha$ such that
in the neighborhood of $\Sigma_2$ we can write the expansion in
$x'^3$ and $t'-t_0$
\begin{eqnarray}\label{coordinate-transform}
t - t_0 & = & _0 \xi^1  x'^3 \ + \ _1 \xi^0 \ (t'-t_0) + ... \ , \nonumber \\
x^3 \equiv  x_3 & = & _0 \xi_3^1 x'^3 \ + \ _1 \xi_3^0 \ (t' - t_0) + ... \ , \\
x^a \equiv  x_a & = & _0 \xi_a^0 + \ _0 \xi_a^1 x'^3 \ + \ _1
\xi_a^0 \ (t' - t_0) + ... \ , \nonumber
\end{eqnarray}
where the expansion coefficients $_n \xi^m$ and $_n \xi_i^m$, with
$n, m = 0, 1, 2, ...$ , are functions of $x'^a$. Notice that this
coordinate transformation is completely general except for the
fact that
\begin{equation}\label{2chi-zero}
_0\xi^0 = \ _0 \xi^0_3 = 0.
\end{equation}

To begin with, we will require that the new coordinates
$\{x'^\alpha\}$ be Gauss coordinates for $V_4$, associated to the
space-like 3-surface $\Sigma'_3$, i.e. to $t'=t_0$. Actually, we
will only require that the $\{x'^\alpha\}$ be Gauss coordinates in
the neighborhood of $\Sigma'_2$, the boundary of $\Sigma'_3$. Reducing our original requirement 
in this way is worth since it is known that Gaussian coordinates, sooner or later,  
develop singularities under appropriate physical conditions
(focussing theorem, see for exemple \cite{Poisson}).

On the other hand, since the equation of the boundary $\Sigma_2$
is $t=t_0$, $x^3 = 0$, this means by definition of boundary that
the metric, $g_{ij}$, and its first derivatives, all them for
$t=t_0$, exist only for, let us say, $x^3
> 0$, at least in some elementary interval around $x^3=0$. Then,
since
\begin{equation} \label{3-metric-prima}
g'_{ij} = - \frac{\partial t}{\partial x'^i} \frac{\partial
t}{\partial x'^j} + \frac{\partial x^k}{\partial x'^i}
\frac{\partial x^l}{\partial x'^j} g_{lk}
\end{equation}
$\Sigma_2$ will still be the boundary of $\Sigma'_3$, provided
that the functions $x^\alpha(x'^\beta)$ and its derivatives, up to
second order included, be well defined coordinates wherever the
metric $g_{ij}$ and its first derivatives are well defined in the
neighborhood of $\Sigma_2$.

Notice that, from Eqs. (\ref{coordinate-transform}), the equation
of $\Sigma_2$ in the new coordinates $\{x'^\alpha\}$ reads $t' =
t_0, x'^3 =0$. Thus, if we name $\Sigma'_2$ the 2-surface $t' =
t_0$, $x'^3 = 0$, we can say that $\Sigma'_2 = \Sigma_2$.

Then, besides requiring that $\{x'^\alpha\}$ be Gauss coordinates
for $V_4$ in the neighborhood of $\Sigma_2$, the boundary of
$\Sigma'_3$, we will require that, according to (\ref{3-metric}), 
\begin{equation}\label{3-metric-prima2}
dl'^2 (t=t_0, x^3=0) \equiv dl'^2|_{\Sigma_2} = f'(x'^a)
\delta_{ij} dx'^i dx'^j.
\end{equation}
Furthermore, we will still require that the new linear and angular
3-momenta, $P'^i$ and $J'^{ij}$ (see (\ref{Pa}), (\ref{P3}) and
(\ref{Jij})), vanish, the last one irrespective of the origin.
That is to say, we want the new coordinate system $\{x'^\alpha\}$
to be an {\em intrinsic} coordinate system as defined in the
Introduction.

From Eq. (\ref{Jij}) we can see very easily that a necessary and
sufficient condition to have  $J^{ij} = 0$, irrespective of the
momentum origin, is that
\begin{equation}\label{int1}
\int \partial_0 g_{3i} \ dx^1 d x^2 = 0,  \quad \forall i,
\end{equation}
which for $i=a$ leads to $P^a = 0$. On the other hand, the three
components of $J^{ij}$ can be more explicitly written
\begin{eqnarray}
J^{12} & = & \kappa \int(x^2 \partial_0 g_{31} - x^1 \partial_0 g_{32}) \ dx^1 d x^2, \label{J12}\\
J^{3a} & = & \kappa \int x^a  \partial_0 g_{33} \  d x^1 d x^2.
\label{J3a}
\end{eqnarray}
Then, aside (\ref{J12}) and (\ref{J3a}) we also have (\ref{int1}).
A sufficient condition to have all this at the same time is that
the $g_{3i}$ metric components be such that
\begin{equation}\label{int2}
\int \partial_0 g_{33} \ dx^1 = \int \partial_0 g_{33} \ d x^2 =
0,
\end{equation}
\begin{equation}\label{int3}
\int \partial_0 g_{3a} \ dx^{(a)} = 0,
\end{equation}
where putting the $a$-index between parenthesis means that the
index is not summed up.

In all: we start from a coordinate system, $\{x^\alpha\}$, where
we have
\begin{equation}\label{G1}
g_{00} = -1, \quad g_{0i} = 0,
\end{equation}
\begin{equation}\label{G2}
g_{3a} (t=t_0)= 0, \quad g_{ij} (t=t_0, x^3=0) = f(x^a)
\delta_{ij},
\end{equation}
and we want to prove that a coordinate transformation
(\ref{coordinate-transform}) exists such that the new components
of the metric satisfy
\begin{equation}\label{G1-prima}
g'_{00} = -1, \quad g'_{0i} = 0,
\end{equation}
\begin{equation}\label{G2-prima}
g'_{ij} (t'=t_0, x'^3=0) = f'(x'^a) \delta_{ij},
\end{equation}
and that, according to (\ref{Pa}), (\ref{P3}), (\ref{int1}),
(\ref{J12}) and (\ref{J3a}), we have:
\begin{equation}\label{int1-prima}
\int \partial'_0 g'_{aa} \ dx'^1 d x'^2 = 0, \ \int \partial'_0
g'_{3i} \ dx'^1 d x'^2 = 0,
\end{equation}
\begin{equation}\label{int2-prima}
\int (x'^2 \partial'_0 g'_{31} - x'^1 \partial'_0 g'_{32}) \ dx'^1
dx'^2= 0,
\end{equation}
\begin{equation}\label{int3-prima}
\int x'^a \partial'_0 g'_{33} \ dx'^1 dx'^2 = 0,
\end{equation}
where $\partial'_0$ means time derivative  with respect the new
time $t'$.

What all these conditions (\ref{G1-prima})-(\ref{int3-prima}) say
about the functions $\ _n \xi^m$ and $\ _n \xi_i^m$ which are
present in the coordinate transformation
(\ref{coordinate-transform})?

In order to answer this question let us first write in the
neighborhood of $\Sigma_2$:
\begin{equation}\label{gij}
g_{ij} = \ _0 g^0_{ij} + \ _0 g^1_{ij} x^3 + \ _1 g^0_{ij} (t -
t_0) + ... ,
\end{equation}
where, according to the notation used in
(\ref{coordinate-transform}), we have:
\begin{equation}\label{gij-coef1}
_0 g^0_{ij} = g_{ij}(t=t_0, x^3=0),
\end{equation}
\begin{equation}\label{gij-coef2}
\ _0 g^1_{ij} = \partial_3 g_{ij} (t=t_0, x^3=0),
\end{equation}
\begin{equation}\label{gij-coef3}
 \ _1 g^0_{ij} = \partial_0 g_{ij} (t=t_0, x^3=0),
\end{equation}
and so on. This means that the expansion coefficients $\ _n
g^m_{ij}$ in (\ref{gij}) are functions only of $x^a$.

Then, Eqs. (\ref{int1-prima}), (\ref{int2-prima}) and
(\ref{int3-prima}) read
\begin{equation}\label{int1-int}
\int \hspace{-2mm}\ _1 g'^0_{aa} \ dx'^1 d x'^2 = 0, \ \int
\hspace{-2mm}\ _1 g'^0_{3i} \ dx'^1 d x'^2 = 0,
\end{equation}
\begin{equation}\label{int2-int}
\int (x'^2 \ _1 g'^0_{31} - x'^1 \ _1 g'^0_{32}) \ dx'^1 dx'^2= 0,
\end{equation}
\begin{equation}\label{int3-int}
\int x'^a \ _1 g'^0_{33} \ dx'^1 dx'^2 = 0,
\end{equation}
where, similarly to (\ref{gij-coef1}), (\ref{gij-coef2}) and
(\ref{gij-coef3}), we have put
\begin{equation}\label{g3a-prima-coef}
\ _1 g'^0_{3a} = \partial'_0 g'_{3a} (t'=t_0, x'^3=0)= \partial'_0
g'_{3a} (t=t_0, x^3=0),
\end{equation}
\begin{equation}\label{g33-prima-coef}
 \ _1 g'^0_{33} = \partial'_0 g'_{33} (t=t_0, x^3=0),
\end{equation}
since, according to (\ref{coordinate-transform}), $t'=t_0, x'^3=0
\Leftrightarrow t=t_0, x^3=0$.

Similarly, Eq. (\ref{G2-prima}) reads now:
\begin{equation}\label{G2-prima-coef}
\ _0g'^0_{ij} = f'(x'^a) \delta_{ij}.
\end{equation}
Thus, with the new notation $\ _n g'^m_{ij}$, the conditions
(\ref{G1-prima})-(\ref{int3-prima}) become (\ref{G1-prima}),
(\ref{int1-int})-(\ref{int3-int}) and (\ref {G2-prima-coef}).

Let us first consider conditions (\ref{G1-prima}). To zero order
in $t'$ and $x'^3$ (that is, strictly on the boundary $\Sigma_2$)
these conditions become
\begin{equation}\label{00-a}
(_1\xi^0)^2 - f (_1\xi^0_3)^2 = 1, \  \ _1\xi^0_a = 0, \ \ _1
\xi^0 \ _0\xi^1 = f \ _1\xi^0_3 \ _0\xi^1_3,
\end{equation}
from $g'_{00} = -1$, $g'_{0a} = 0$ and $g'_{03} = 0$,
respectively.

On the other hand, conditions (\ref {G2-prima-coef}) become
\begin{equation}\label{00-b}
f' \delta_{ab} = f \delta_{cd} \, \frac{\partial
_0\xi^0_c}{\partial x'^a} \, \frac{\partial _0\xi^0_d}{\partial
x'^b}, \ \ _0\xi^1_a = 0, \ f(_0\xi^1_3)^2 - (_0\xi^1)^2 = f',
\end{equation}
from $_0g'^0_{ab} = f' \delta_{ab}$, $_0g'^0_{3a} = 0$ and
$_0g'^0_{33} = f'$, respectively.

It can be seen that the general solution of the system
(\ref{00-a}) and (\ref{00-b}) is
\begin{equation}\label{s1}
_1\xi^0_a = \ _0\xi^1_a = 0.
\end{equation}
\begin{equation}\label{s2}
_1 \xi^0 = \sqrt{\frac{f}{f'}} \, \, _0 \xi^1_3 = \cosh \psi,
\end{equation}
\begin{equation}\label{s3}
\frac{1}{\sqrt{f'}}\, \,  _0 \xi^1 = \ \sqrt{f} \, _1 \xi^0_3 =
\sinh \psi,
\end{equation}
plus
\begin{equation}\label{Jacobiana}
M_{ab} \equiv \frac{\partial _0\xi^0_a}{\partial x'^b} = \lambda
\, \left(
\begin{array}{cc}
\cos \theta & \sin \theta \\
- \sin \theta & \cos \theta
\end{array} \right),
\,  \lambda \equiv \sqrt{f'/f},
\end{equation}
the Jacobian matrix of the conformal transformation in two
dimensions. In (\ref{s2}), (\ref{s3}) and (\ref{Jacobiana}) the
functions $\psi$, $\lambda$ and $\theta$ are arbitrary functions
of $x'^a$. Notice that (\ref{Jacobiana}) says that in the
integrals (\ref{int1-int})-(\ref{int3-int}) we can put $dx'^1
dx'^2 = \lambda^{-2} dx^1 dx^2$.

We still must have:
\begin{eqnarray}
_1g'^0_{3a} & = &  (f \ _1 \xi^1_b  + \ _1g^0_{3b} \ _1 \xi^0 \ _0
\xi^1_3)M_{ba} + \ f \ _0 \xi^1_3 \ _1 \xi^0_{3,a} - \ _0 \xi^1 \ _1 \xi^0_{,a} \, ,\label{10a} \\
_1g'^0_{33} & = & 2 ( f \ _0 \xi^1_3 \ _1 \xi^1_3 - \ _0 \xi^1 \
_1 \xi^1) + \ _1g^0_{33} \ _1 \xi^0 (_0 \xi^1_3)^2  \, , \label{10b} \\
_1g'^0_{aa} & = & (_1g^0_{bc} \ _1 \xi^0 + \ _0g^1_{bc} \ _1
\xi^0_3) M_{ba} M_{ca} = \lambda^2 (_1g^0_{aa} \ _1 \xi^0 + \
_0g^1_{aa} \ _1 \xi^0_3) \, , \label{10c}
\end{eqnarray}
where $_1g'^0_{3a}$, $_1g'^0_{33}$ and $_1g'^0_{aa}$ are functions
of $x'^a$ such that (\ref{int1-int}), (\ref{int2-int}) and
(\ref{int3-int}) are satisfied. The derivative with respect $x'^a$
is denoted by $,a$ (for instance, $\ _1 \xi^0_{3,a} \equiv
\frac{\partial \ _1 \xi^0_3}{
\partial x'^a}$).

In Eqs. (\ref{10a}) and (\ref{10b}) new expansion coefficients $_1
\xi^1_i $ and $_1 \xi^1$ appear, which are not included in
(\ref{s1})-(\ref{Jacobiana}). But they appear in Eq.
(\ref{G1-prima}) when it is taken to zero order in $t'$ and order
one in $x'^3$ (remember that up to now we have only considered the
lowest order of this equation), which becomes:
\begin{eqnarray}
_0g'^1_{0a} & = & (f \ _1 \xi^1_b + \ _1g^0_{3b} \ _0 \xi^1  \ _1
\xi^0_3) M_{ba} + f \ _1 \xi^0_3 \ _0 \xi^1_{3,a} - \ _1 \xi^0 \
_0 \xi^1_{,a} = 0,  \label{01a} \\
_0g'^1_{03} & = &  f (_1 \xi^0_3 \  _0 \xi^2_3 + \ _1 \xi^1_3 \ _0
\xi^1_3) - \ _1 \xi^0 \  _0 \xi^2 -  \ _1 \xi^1 \ _0 \xi^1 +  \
_1g^0_{33} \ _0 \xi^1 \  _1 \xi^0_3 \ _0 \xi^1_3 = 0, \qquad
\label{01b}\\
_0g'^1_{00} & = &  2 ( f \ _1 \xi^0_3 \ _1 \xi^1_3 - \ _1 \xi^0 \
_1 \xi^1) +  \ _1g^0_{33} \ _0 \xi^1  (_1 \xi^0_3)^2 = 0.
\label{01c}
\end{eqnarray}
Therefore, we must fit the new expansion coefficients, $_1 \xi^1_i
$ and $_1 \xi^1$, plus the arbitrary functions $\lambda$,
$\theta$, and $\psi$, of Eqs. (\ref{s2})-(\ref{Jacobiana}), in
order to satisfy the system (\ref{10a})-(\ref{10c}) plus
(\ref{01a})-(\ref{01c}). Let us show that this can always be done.

First, since the Jacobian matrix $M_{ab}$ is regular, we can
always fit the $_1 \xi^1_b$ such that the two Eqs. (\ref{10a})  be
satisfied. Second, since $f \neq 0$, ($dl^2$ is strictly positive)
and (see Eq. (\ref{s2})) $_0 \xi^1_3 \neq 0$, we can fit $_1
\xi^1_3$ such that Eq. (\ref{10b}) be satisfied too. Furthermore,
it can be seen (see Appendix \ref{ap-A}) that, to get $P'^3 = 0$,  $\psi$ can always be
fitted such that Eq. (\ref{10c}) becomes satisfied.

Next, we consider the three remaining Eqs.
(\ref{01a})-(\ref{01b}). Since (see again (\ref{s2})) $_1 \xi^0
\neq 0$ we can fit $_0 \xi^2$ such as to have (\ref{01b}).
Similarly for Eq. (\ref{01c}) by fitting $_1 \xi^1$. Finally, it
can be proved (see Appendix \ref{ap-B}) that the Jacobian matrix
(\ref{Jacobiana}) can always be fitted in order to have Eq.
(\ref{01a}) satisfied.

In all, we have just proved that for any {\em universe} there
always exist intrinsic coordinate systems, that is Gaussian
coordinates, $\{x'^\alpha\}$,  satisfying the  supplementary
conditions (\ref{G2-prima-coef}), and such that $P'^i = 0$ and,
irrespective of the angular momentum origin, $J'^{ij} = 0$.

\section{Creatable universes}
\label{sec-4}

Let it be a {\em universe} that we have referred to intrinsic
coordinates $\{x'^\alpha\}$. Then, we will call that {\em
universe} a {\em creatable universe} if in these coordinates we
also have:
\begin{equation}\label{creatable}
P'^0 = 0, \quad J'^{0i} = 0.
\end{equation}
This means, according to Eqs. (\ref{P0}), (\ref{J0a}) and
(\ref{J03}), that
\begin{eqnarray}
P'^0 & = & - \kappa \int \hspace{-2mm}\ _0g'^1_{aa} d x'^1 d x'^2 = 0 , \label{P0-prima}\\
J'^{0a} & = &  \kappa \int  x'^a  \ _0 g'^1_{bb} d x'^1 d x'^2 = 0 , \label{J0a-prima}\\
J'^{03} & = & - \kappa \int \hspace{-2mm}\ _0g'^0_{aa} d x'^1 d x'^2
= - 2 \kappa \int f' d x'^1 d x'^2 = 0. \qquad \label{J03-prima}
\end{eqnarray}
that is, $_0g'^1_{aa}$ and $f'$ must be such that the above four
integrals vanish.

On the other hand, we find after some calculation
\begin{equation}
_0g'^1_{aa} = \ (_1g^0_{bc} \ _0 \xi^1 + \ _0g^1_{bc} \ _0
\xi^1_3) M_{ba} M_{ca} = \lambda^2 (_1g^0_{aa} \ _0 \xi^1 + \
_0g^1_{aa} \ _0 \xi^1_3) \label{01d}
\end{equation}
which can be compared with (\ref{10c}). Notice that here we are
left with no more freedom to fit a given value of $_0g'^1_{aa}$ in
order to have (\ref{P0-prima}) and (\ref{J0a-prima}): in fact,
both, the Jacobian matrix $M_{ab}$, plus  $_0 \xi^1$ and $_0
\xi^1_3$ (that is to say, plus $\psi$, according to (\ref{s2}) and
(\ref{s3})), have already been fitted such as to have intrinsic
coordinates. This means, that: 

A {\em universe} is not necessarily a creatable universe, which even if expected is a very remarkable result.

Now, before we can continue, we must say something about Eq.
(\ref{J03-prima}), that would have to be satisfied if, according
to our definition, we have a creatable universe. Since $f'$ is
strictly positive it seems at first sight that (\ref{J03-prima})
can only be satisfied in any one of the two following cases:
first, if the area of $\Sigma_2$ vanishes (in which case $f'$ should
remain conveniently bounded when we approach $\Sigma_2$; notice
that the boundary $\Sigma_2$ could not belong to $\Sigma'_3$, in
which case $f'$ could go to infinite when we approach $\Sigma_2$);
second, if $f'$ goes to zero when we approach $\Sigma'_2$, which
means again that $\Sigma_2$ does not belong to $\Sigma'_3$.

But, actually, these are not the only cases where we can have
(\ref{J03-prima}), since, according to what is said at the end of
section \ref{sec-2}, $\Sigma_2$ could have several different
sheets, and it could happen that the different contributions from
these different sheets compensate among them to give a vanishing
value for $\int f' dx'^1 dx'^2$. Thus, in Minkowski space, $M_4$, in
Lorentzian coordinates (which are {\em intrinsic} coordinates) we
have $f' = 1$. But, $\Sigma_2$ is made from six sheets, the six
faces of a cube that increases without limit. Then, the two
contributions corresponding to two opposite faces cancel each one
to the other.

Anywise, some one could argue that we could only define a given
{\em universe} as a creatable universe if $P^\alpha = J^{\alpha
\beta} = 0$ for ANY intrinsic coordinate system. But this would be
an exceeding demand since not even the case of the Minkowski
space-time, $M_4$,  would satisfy such a strong requirement.
Actually, one type of intrinsic coordinates for this {\em
universe} are the standard Lorentz coordinates. Furthermore, in
these coordinates, all 4-momenta, $P^\alpha$ and $J^{\alpha
\beta}$ vanish, so that this {\em universe} is a creatable
universe according to the definition we have just given.
Nevertheless, it can be easy seen (see Appendix \ref{ap-C}) that
starting  from Lorentz coordinates, one can always make an
elementary coordinate transformation leading to new, non
Lorentzian, intrinsic coordinates, such that the new energy $P'^0$
does no more vanish. Obviously, according to section \ref{sec-3},
this elementary coordinate transformation has to be one where the
infinitesimal version of the coefficients $_0 \xi^0$ and $_0
\xi^0_3$ do not vanish, that is Eq. (\ref{2chi-zero}) does not
more occur.

The reason for this non vanishing energy, $P'^0$, in $M_4$ is
that, by doing the above elementary coordinate transformation, we
have left a coordinate system (the Lorentzian one) which was well
adapted to the symmetries of the Minkowskian metric: the ones tied
to the ten parameters of the Poincar\'e group.

Thus, given a {\em universe} which has $P^\alpha = J^{\alpha
\beta} = 0$ for some intrinsic coordinate system, if there are
other intrinsic coordinates where this vanishing is not preserved,
we should consider that this non preservation expresses the fact
that the new intrinsic coordinates are not well adapted to some
basic metric symmetries. To which symmetries, to be more precise?
In general terms, to the ones which allow us to have just
vanishing linear and angular 4-momenta for some intrinsic
coordinate system.

In other words: in spite of the apparent freedom in the choice of the coordinate frame, 
we have characterized in an intrinsic way if a {\it universe} has or has not vanishing $4$-momenta.
In our framework, in order to have this vanishing, we only need to find ONE {\it intrinsic} coordinate frame
where  $P^\alpha$ and $J^{\alpha \beta}$ vanish, which, as we have just explained, can only been found in some special {\it universes}.

\section{The perturbed FLRW universes}
\label{sec-5}

In Ref. \cite{Lapiedra-Saez} the creatibility of perturbed FLRW
universes was addressed. The main result of that paper which
concerns us here is that in the flat case it is found that the
energy is infinite, $P^0 = \infty$, for inflationary scalar
perturbations plus arbitrary tensor perturbations. This seems to
say that inflationary perturbed flat FLRW universes are not
creatable. Nevertheless, as it has been already stressed at the end of the Introduction, 
this assessment needs to be validated in
the new framework we have developed in the present paper, where
creatibility can only be considered for intrinsic coordinate
systems, i. e., systems where, in particular, the linear and
angular 3-momenta, $P^i$ and $J^{ij}$, vanish.

Then, we prove next that both momenta vanish in the coordinate
system where it was obtained that $P^0 = \infty$.  Therefore, we
conclude that, in the new framework of the present paper, the non
creatibility of the inflationary perturbed flat FLRW universe
remains unchanged.

Let us prove first that $P^i$ vanish. According to Ref.
\cite{Lapiedra-Saez} we write the perturbed 3-space metric $dl^2$
as
\begin{equation}\label{per-metric1}
dl^2 = \frac{a^2(t)}{(1 + \frac{{\rm k}}{4} r^2)^2}(\delta_{ij} +
h_{ij}) dx^i d x^j,
\end{equation}
where $a(t)$ is the cosmic expansion factor.

In the flat case, ${\rm k}=0$, when considering inflationary
scalar perturbations, the perturbed 3-space metric, $h_{ij}$,
reads
\begin{equation}\label{per-metricx}
h_{ij}(\vec{x}, \tau) = \int \exp(i \vec{k} \cdot \vec{x})
h_{ij}(\vec{k}, \tau)  d^3 k
\end{equation}
with the following expression for the Fourier transformed function
$h_{ij}(\vec{k}, \tau)$:
\begin{equation}\label{per-metrick}
h_{ij}(\vec{k}, \tau) = h(\vec{k}, \tau) \hat{k}_i \hat{k}_j + 6
\eta (\vec{x}, \tau) (\hat{k}_i \hat{k}_j - \frac{1}{3}
\delta_{ij}).
\end{equation}
Here $h \equiv h_{kk}$ and $\eta$ are convenient functions,
$\hat{k}_i \equiv k_i/k$, $k \equiv \sqrt{k_i k^i}$, and $\tau$ is
defined such that $dt/d\tau \equiv a$.

According to Eq. (\ref{linear momentum}):
\begin{equation}\label{limP}
P^i = \lim_{r \to \infty} \frac{r^2}{16 \pi G} \int I^i d^3 k
\end{equation}
where
\begin{eqnarray}\label{intI}
I^i & \equiv & \int  \exp(i \vec{k} \cdot \vec{x})[\dot{h}_{kk}(\vec{k}, \tau) \delta_{ij} -
\dot{h}_{ij}(\vec{k}, \tau)] n_j  d \Omega \nonumber\\
& = & \int  \exp(i \vec{k} \cdot \vec{x})[\dot{h}(\vec{k}, \tau)
(\delta_{ij} - \hat{k}_i \hat{k}_j) + 6 \dot{\eta} (\vec{x}, \tau)
(\frac{1}{3} \delta_{ij} - \hat{k}_i \hat{k}_j)] n_j  d \Omega.
\end{eqnarray}
Here,  the dot stands for the time, $t$, derivative and with $d
\Omega$ the integration element of solid angle.

Notice that here we have taken as $\Sigma_2$ the $2$-surface
$t=t_0$, $r = R \to \infty$, instead of the six faces of the over
growing cube reported at the end of section \ref{sec-2}. Of
course, the $3$-volume element remains $dx^1 dx^2 dx^3$ since
these are the corresponding intrinsic coordinates. We can take $r
= R \to \infty$ for $\Sigma_2$ because of the choice of the above
cube had only the function of making easier the proof of the
existence of intrinsic coordinates.

On the other hand, one easily finds
\begin{equation}\label{intPhi}
\int  \exp(i \vec{k} \cdot \vec{x}) n_i d \Omega = \frac{4 \pi
i}{k r}(\frac{\sin kr}{kr} - \cos kr) \hat{k}_i \equiv \Phi (k, r)
\hat{k}_i
\end{equation}
where what is important for us here is that $\Phi$ does not depend
on $\hat{k}_i$. Then
\begin{equation}\label{Ii}
I^i =  \Phi [6 \dot{\eta} (\vec{x}, \tau) (\frac{1}{3} \hat{k}_i -
\hat{k}_i)] = - 4 \Phi \dot{\eta} (\vec{x}, \tau) \hat{k}_i.
\end{equation}

But, as it has been quoted in \cite{Lapiedra-Saez}, in the
case of inflationary scalar perturbations, in which we are
interested here, $\eta(\vec{k}, \tau)$ does not actually depend on
$\hat{\vec{k}}$. Then, by symmetry, $\int I^i d^3 k = 0$, and so,
$P^i =0$ for any time.

Next, we consider general tensor perturbations and we see that
$P^i$ vanish too. As quoted again in Ref. \cite{Lapiedra-Saez},
the above Fourier transformed function $h_{ij}(\vec{k}, \tau)$
reads now:
\begin{equation}\label{per-tensor}
h_{ij}(\vec{k}, \tau) = H(k, \tau) \epsilon_{ij}(\hat{k}),
\end{equation}
where the symmetric matrix $\epsilon_{ij}$ is transverse and
traceless:
\begin{equation}\label{ttless}
\epsilon_{ij} k_i = 0, \quad \epsilon_{ii} = 0.
\end{equation}

The above $I^i$ integral becomes now
\begin{equation}\label{intIten}
I^i = - \int \exp(i \vec{k} \cdot \vec{x}) H(k, \tau)
\epsilon_{ij} n_j d \Omega,
\end{equation}
which according to (\ref{intPhi}) and the first equation in
(\ref{ttless}) becomes $I^i = 0$. This is, we have again $P^i =
0$.

Thus, when inflationary scalar and general tensor perturbations
are both present we have $P^i = 0$, as we wanted to prove.

The next step will be to prove that, for any time, $J^{jk}$ vanish
too for both types of perturbations. Let us first consider
inflationary scalar perturbations, that is, Eq.
(\ref{per-metrick}).

According to Eq. (\ref{Jij}):
\begin{equation}\label{J-per}
J^{jk} = \lim_{r \to \infty} \frac{r^3}{16 \pi G} \int  I^{jk} d^3
k,
\end{equation}
where
\begin{equation}\label{Ijk}
I^{jk} = \int  \exp(i \vec{k} \cdot \vec{x})[n_k \dot{h}_{ij}
(\vec{k}, \tau) - n_j \dot{h}_{ki}(\vec{k}, \tau)] n_i d \Omega.
\end{equation}

But, obviously:
\begin{equation}\label{intnn}
\int \exp(i \vec{k} \cdot \vec{x}) n_i n_j  d \Omega \propto
\delta_{ij}, \, \,  \hat{k}_i \hat{k}_j,
\end{equation}
that is, the calculation of this integral must give a contribution
which goes like $\delta_{ij}$, and another one which goes like
$\hat{k}_i \hat{k}_j$. Then, it is easy to verify that when these
two kinds of contributions are introduced in (\ref{Ijk}) we obtain
identically $I^{jk} = 0$, and so $J^{jk} = 0$.

Finally, we will consider general tensor perturbations, that is,
$h_{ij}(\vec{k}, \tau)$ given by Eqs. (\ref{per-tensor}) and
(\ref{ttless}). In this case (\ref{Ijk}) becomes
\begin{equation}\label{Ijk2}
I^{jk} = \dot{H}(k, \tau) \int  \exp(i \vec{k} \cdot \vec{x}) (n_k
\epsilon_{ij} - n_j \epsilon_{ki}) n_i  d \Omega.
\end{equation}

But having in mind (\ref{intnn}) and the first equation of
(\ref{ttless}) it is straightforward to see that $I^{jk}$ and then
$J^{jk}$ vanish.

All in all, for any time, $P^i$ and $J^{ij}$ vanish in the same
coordinate system where it was proved (see Ref.
\cite{Lapiedra-Saez}) that $P^0 = + \infty$. Then, we can assert
that our perturbed flat FLRW universe is really a non creatable one.

On the other hand, it can be easily seen that in the present new
approach, as in \cite{Lapiedra-Saez}, perturbed closed FLRW
universes are creatable, while perturbed open FLRW universes are
not.

\section{Final considerations}
\label{sec-6}

The energy of Friedmann-Lema{\^{\i}}tre-Robertson-Walker (FLRW) cosmologies has been calculated by different authors using divers procedures, like pseudotensorial methods based on specific choices of coordinates \cite{Rosen,Garecki-1995,Garecki-2008,Mitra,Berman}, or Hamiltonian methods imposing boundary conditions \cite{Chen-Liu-Nester-2007,Nester-So-Va-08}, or by choosing an appropriate background configuration \cite{Katz-Bi-LyBell-97,BibbonaFF}, or even by other procedures \cite{Cooperstook}. Quasi-local approaches have also been  extensively considered, providing distinct results because of the different used definitions \cite{Hayward,Afshar-2009}.  Many authors (us including) agree with the following statement:  the total energy vanishes both for closed and flat FLRW universes,  but diverges to $- \infty$ for the class of open models (negative curvature index,  $k=-1$). Thus, the closed and flat FLRW universes would be creatable, but the open one would be not.

However, there is no full agreement in the current literature on these energy values (cf \cite{Garecki-2008,Faraoni-Cooperstock-2003}) although, in our opinion,  their goodness becomes supported by the rightness of the criteria we have implemented in the present paper to define proper values for all 
4-momenta components. Among the references which agree with these FLRW values are \cite{Cooperstook,Rosen}

Let us specify that these values were obtained by us in \cite{Ferrando}, 
but the translation of the result from the old framework in \cite{Ferrando} to the new one in the present paper is straightforward.

Notice that the same conclusion follows from the results obtained
in \cite{Katz-Bi-LyBell-97} concerning integral conservation laws
with respect to a given background and its associated isometry
group, but only when this background is the flat space-time. 

On the other hand, the creatibility of the perturbed FLRW universes (see Section \ref{sec-5}) should be also
analyzed following the approach of Ref. \cite{Katz-Bi-LyBell-97}.
In this case, the above conclusion about the non-perturbed case
strongly suggests that the results presented in Sect. 
\ref{sec-5} could be recovered from the results of
\cite{Katz-Bi-LyBell-97} under these assumptions: (i) the
considered background is the Minkowski space-time, (ii) the
conservation laws are referred to the background isometries,  and
(iii) the perturbed metric and the energy content are considered
in some synchronous gauge (by taking Gauss coordinates).

Now, before ending the paper we would like to point out that the
main interest of it could be to give a criterion to discard from
the very beginning as much as possible space-times as candidates
to represent our actual Universe. The criterion could be that good
initial candidates must be {\it creatable} universes. Thus, as commented above, 
in \cite{Lapiedra-Saez} it was claimed that, within the inflationary
perturbed FLRW universes, only the closed case corresponds to a
{\it creatable} universe. Of course, the criterion is not a consequence of the theory of the General Relativity when  applied to cosmology. It is only a guess, one appearing in the literature at last since 1973 \cite{Albrow,Tryon} that we find so appealing, in our opinion, as to deserve that its consequences be explored, as we have continued to do in the present paper. This result, obtained in   \cite{Lapiedra-Saez} in a non conclusive way, has been fully validated
in the framework of the present paper, as it has been proved in Sect. \ref{sec-5}.
Similarly, since some other space-times have lately been
considered as candidates to represent our Universe (see for
example, \cite{LRey-Luminet}, \cite{Jaffe}), we could check them
to see if they fulfill the above criterion of creatibility. When
making this checking, in the case we obtained $P^\alpha = 0$ and
$J^{\alpha \beta} = 0$ for a given $t = t_0$, we still had to
verify that the result does not depend of the value of $t_0$, that
is, we would have to verify {\em a posteriori} that we were
dealing with a space-time which is a {\em universe}. All this
would deserve some future work. \\

{\bf Acknowledgements} This work has been supported by the Spanish Ministerio de
Ciencia e Innovaci\'on MICINN-FEDER project No. FIS2009-07705.


\appendix

\section{Fitting the function $\psi$ to get $P'^3=0$}
\label{ap-A}

We must fit $\psi$ such that $_1g'^0_{aa}$, given by (see
(\ref{10c}))
\begin{equation}
_1g'^0_{aa}  = \lambda^2 (_1g^0_{aa} \ _1 \xi^0 + \ _0g^1_{aa} \
_1 \xi^0_3) \, , \label{A1}
\end{equation}
gives $P'^3 = 0$. Notice that according to Eq. (\ref{P3}) we have
\begin{equation}\label{A2}
P'^3 = \kappa \int \hspace{-2mm}\ _1 g'^0_{aa} \ d x'^1 d x'^2 \,.
\end{equation}
On the other hand, from (\ref{s2}) and (\ref{s3}), the equation
(\ref{A1}) can be written as
\begin{equation}\label{A3}
a = b \cosh \psi + c \sinh \psi \, ,
\end{equation}
where
\begin{equation}\label{A4}
a \equiv \ _1 g'^0_{aa}, \quad b \equiv \lambda^2 \ _1 g^0_{aa},
\quad c \equiv \frac{\lambda^2}{\sqrt f} \ _0g^1_{aa}
\end{equation}
Then, putting $\cosh \psi \equiv x$, we obtain the algebraic
second order equation
\begin{equation}\label{A5}
(b^2 -c^2) x^2 -2 a b x + a^2 + c^2 = 0 \, ,
\end{equation}
that only has real solutions if
\begin{equation}\label{A6}
a^2 + c^2 \geq b^2 \, .
\end{equation}
But we can ensure it by taking $a$ large enough. This can always
be made since if $a \equiv \hspace{-2mm}\ _1 g'^0_{aa} \neq 0$ is
such that $\int a \ dx'^1 d x'^2 = 0$, then we also will have
$\int K a dx'^1 d x'^2 = 0$, with $K$ a constant whose absolute
value, $|K|$, is as large as
we wanted.%
\footnote{The singular case $a \equiv \ _1 g'^0_{aa} = 0$, would
give as a solution for (\ref{A3}) $\tanh \psi = - b/c$, which only
exists if $|b/c|<1$. \label{fn3}}
Furthermore, if $|K|$ is large enough, we can easily see that for
the new coefficient $a$, that is, for $K a$, one at least of the
$x$ solutions is larger than one, as it must be.
\section{Fitting conveniently the functions $\lambda$ and $\theta$
or the functions $\lambda$ and $\psi$} \label{ap-B}

According to what is said at the end of section \ref{sec-3}, we
must fit the functions $\lambda$ and $\theta$ such that Eq.
(\ref{01a}) be satisfied. Taking in account (\ref{10a}), the  Eq.
(\ref{01a}) becomes:
\begin{equation}\label{B1}
_1g'^0_{3a} =  (\ _1 \xi^0 \ _0 \xi^1_3 - \ _0 \xi^1  \ _1
\xi^0_3) \ _1g^0_{3b} M_{ba} + \ f ( \ _0 \xi^1_3 \ _1 \xi^0_{3,a}
- \ _1 \xi^0_3 \ _0 \xi^1_{3,a}) + \ _1 \xi^0 \ _0 \xi^1_{,a} - \
_0 \xi^1 \ _1 \xi^0_{,a} \quad
\end{equation}
where $\ _1 \xi^0_{3,a} \equiv \frac{\partial \ _1 \xi^0_3}{
\partial x'^a}$, and so on. Furthermore, having in mind (\ref{s2}),
(\ref{s3}) and the definition of $\lambda$ in (\ref{Jacobiana}),
Eq. (\ref{B1}) becomes:
\begin{equation}\label{B2}
\ _1g'^0_{3a} = \lambda (M_{ba} \ _1 g^0_{3b} + X_a),
\end{equation}
where we have put
\begin{equation}\label{B3}
X_a \equiv \frac{2}{\sqrt f} \ \frac{\partial \psi}{\partial x'^a}
.
\end{equation}
Then, from (\ref{Jacobiana}), we obtain the system
\begin{eqnarray}
\lambda^2 (\ _1 g^0_{31} \cos \theta -
\ _1 g^0_{32}  \sin \theta)  & = & - \lambda X_1 + \ _1g'^0_{31} \label{B4}\\
\lambda^2 (\ _1 g^0_{32} \cos \theta + \ _1 g^0_{31} \sin \theta)
& = &  - \lambda X_2 + \ _1g'^0_{32}.\label{B5}
\end{eqnarray}
Notice that, in this system, the functions $\ _1 g'^0_{3a}$ are
defined modulus an arbitrary constant factor $K$ (as it was,
above, the case with $\ _1 g'^0_{aa}$). This means that, in
(\ref{B4}) and (\ref{B5}), we can take $\ _1 g'^0_{3a}$ as small
as we want, provided that the original $\ _1 g'^0_{3a}$ remain
bounded (the unbounded special case will be considered next),
which in turn means that we can take as the system to solve
\begin{eqnarray}
\lambda(\ _1 g^0_{31} \cos \theta -
\ _1 g^0_{32}  \sin \theta)  & = & - X_1 \label{B6}\\
\lambda (\ _1 g^0_{32} \cos \theta + \ _1 g^0_{31} \sin \theta)  &
= &  - X_2, \label{B7}
\end{eqnarray}
whose unique solution, out of the singular case $\ _1 g^0_{3a} =
0$,  is
\begin{equation}\label{B8}
\lambda \cos \theta = - \frac{\ _1 g^0_{31} X_1 + \ _1 g^0_{32}
X_2}{(\ _1 g^0_{31}) ^2 + (\ _1 g^0_{32})^2} \equiv Y_1,
\end{equation}
\begin{equation}\label{B9}
 \lambda \sin \theta = \frac{\ _1 g^0_{32} X_1 - \ _1 g^0_{31} X_2}{(\ _1
g^0_{31}) ^2 + (\ _1 g^0_{32})^2} \equiv Y_2,
\end{equation}
that is to say
\begin{equation}\label{B10}
\lambda = \sqrt{Y_1^2 + Y_2^2}, \quad  \tan \theta =
\frac{Y_2}{Y_1}.
\end{equation}

To complete the above discussion let us consider the special case
where $\ _1 g'^0_{3a}$ goes to infinite when we approach
$\Sigma_2$. (Obviously this will have to be compatible with the
vanishing of the integrals $\int \hspace{-2mm}\ _1 g'^0_{3a} dx'^1
dx'^2$). In this case, the system (\ref{B4}), (\ref{B5}),
becomes:
\begin{eqnarray}
\lambda^2 (\ _1 g^0_{31} \cos \theta -
\ _1 g^0_{32}  \sin \theta)  & = & \ _1g'^0_{31} \label{B11}\\
\lambda^2 (\ _1 g^0_{32} \cos \theta + \ _1 g^0_{31} \sin \theta)
& = &  \ _1g'^0_{32},  \label{B12}
\end{eqnarray}
with $\ _1g'^0_{3a}$ going to infinite, whose solution is
\begin{equation} \label{B13}
\lambda^2 = \infty,  \tan \theta = \lim_{\ _1 g'^0_{3a}\rightarrow
\infty} \frac{\ _1 g^0_{31} \ _1 g'^0_{32}- \ _1 g^0_{32} \ _1
g'^0_{31}}{\ _1 g^0_{31} \ _1 g'^0_{31} + \ _1 g^0_{32} \ _1
g'^0_{32}}.
\end{equation}

We could still consider the remaining two special cases where,
only one of the two functions $\ _1 g'^0_{3a}$ goes to infinite,
but the reader can see easily than also in both cases a solution
exists for $\lambda$, $\theta$.

To end with this Appendix \ref{ap-B}, let us consider the above
singular case $\ _1 g^0_{3a} = 0$. It seems that now the four Eqs.
(\ref{10a}) and (\ref{01a}) cannot always be satisfied by fitting
$\ _1 \xi^1_b$ and $M_{ab}$ since these four unknown functions
appear now through only two quantities $\ _1 \xi^1_b M_{ba}$.

Nevertheless, let us proceed along the following lines:

As far as Eq. (\ref{01a}) is concerned, we always can satisfy it
by fitting some convenient values of $\ _1 \xi^1_b$, since $f \neq
0$ and $M_{ab}$ is a regular matrix.

On the other hand, according to (\ref{B2}) and (\ref{B3}), Eq.
(\ref{10a}) reads now
\begin{equation}\label{B14}
\ _1g_{3a}'^0 = \frac{2 \lambda }{\sqrt f} \ \frac{\partial
\psi}{\partial x'^a} .
\end{equation}

Using $\lambda$ as an integrating factor, we always can find a
family of solutions $\psi$ of these two equations. Then, we must
fit this family of solutions such that the Eq. (\ref{10c}) we are
left with,
\begin{eqnarray}\label{B15}
\ _1g_{aa}'^0 & = & \lambda^2( \ _1g_{aa}^0 \cosh \psi + \frac{\
_0g_{aa}^1}{\sqrt{f}} \sinh \psi),
\end{eqnarray}
becomes satisfied. To see that this is also possible, in
(\ref{B14}) we will choose $\ _1g'^0_{3a}= \epsilon_a g_3$, with
$\epsilon_a = 1$, $\forall a$, and $g_3$ a function such that
$\int g_3 dx'^1 dx'^2 =0$. In this case we have $\frac{\partial
\psi}{\partial x'^1}=\frac{\partial \psi}{\partial x'^2}$, that is
$\psi$ is a function of $x'^1 + x'^2 \equiv y_1$, but not of  $y_2
\equiv x'^1 - x'^2$:
\begin{equation}\label{B16}
\frac{\partial \psi}{\partial y_2} = 0.
\end{equation}
Then, let us integrate (\ref{B15}) along $y_2$ over $\Sigma_2$. We
will have
\begin{equation}\label{B17}
a = b \cosh \psi + c \sinh \psi \, ,
\end{equation}
with
\begin{equation}\label{B18}
a = \int \hspace{-2mm}\ _1g'^0_{aa}  d y_2, \, b = \int \lambda^2
\ _1g^0_{aa} d y_2, \, c = \int \frac{\lambda^2}{\sqrt f} \
_0g^1_{aa} d y_2,
\end{equation}
where, like $\psi$,  the coefficients $a$,  $b$, $c$,  depend only
on  $y_1$.  On the ground of what was said for the coefficient $a$
of Appendix \ref{ap-A}, the present coefficient $a$ is also as
greater as we want. Then, we can conclude that (\ref{B17}) always
have a solution for $\psi$ for any function $\ _1g'^0_{aa}$ such
that $\int \hspace{-2mm}\ _1g'^0_{aa} dx'^1 dx'^2 = 0$. That is to
say, Eqs. (\ref{10a}), (\ref{10c}) and (\ref{01a}) can all be
satisfied at the same time, as we wanted to prove in the present
singular case $\ _1 g^0_{3a} = 0$.
\section{The counter example of Minkowski space}
\label{ap-C}

In section \ref{sec-4}, we claim that if we have a {\em universe}
such that its ten 4-momenta vanish for some given intrinsic system
of coordinates, we cannot  hope to keep this ten-fold vanishing
against any coordinate change going to new intrinsic coordinates.
The reason of this is that even Minkowski space, $M_4$, have not
such a property.

In order to see this, refer $M_4$ to Lorentzian coordinates. These
are obviously intrinsic coordinates, in the sense of the present
paper. Furthermore, all ten 4-momenta vanish in this Lorentzian
frame. Thus, according to our definition, $M_4$ is an example of
creatable universe. Then, let us make some general infinitesimal
coordinate transformation:
\begin{equation}\label{C1}
x^\alpha = x'^\alpha + \epsilon^\alpha(x)\, ,
\end{equation}
where the old coordinates, $\{x^\alpha\}$, are Lorentzian
coordinates. Let us subject the functions $\epsilon(x)$ to the
condition that the new coordinates $\{x'^\alpha\}$ be intrinsic
coordinates. That is, the new metric components
\begin{equation}\label{C2}
g'_{\alpha\beta} = \eta_{\alpha\beta} + \eta_{\alpha\rho}
\partial_\beta \epsilon^\rho + \eta_{\beta\rho}
\partial_\alpha \epsilon^\rho
\end{equation}
has to satisfy on the one hand, Eqs. (\ref{G1-prima}) and
(\ref{G2-prima}) (the first one up to zero order in $t'- t_0$ and
order one in $x'^3$). On the other hand, the time derivatives
$\partial'_0 g'_{3i}$, $\partial'_0 g'_{aa}$, must fulfill the
conditions (\ref{int1-int})-(\ref{int3-int})
\begin{equation}\label{C3}
\int \hspace{-2mm}\ _1 g'^0_{aa} \ dx^1 d x^2 = 0, \ \int
\hspace{-2mm}\ _1 g'^0_{3i} \ dx^1 d x^2 = 0,
\end{equation}
\begin{equation}\label{C4}
\int (x'^2 \ _1 g'^0_{31} - x'^1 \ _1 g'^0_{32}) \ dx^1 dx^2= 0,
\end{equation}
\begin{equation}\label{C5}
\int x'^a \ _1 g'^0_{33} \ dx^1 dx^2 = 0,
\end{equation}
which mean that $P'^i = 0$ and that, irrespective of the origin of
the angular momentum, $J'^{ij} = 0$ (notice that to first order we
can put $dx^1 dx^2$ instead of $dx'^1 dx'^2$).

After some elementary calculations, all these conditions are
written:
\begin{equation}\label{C6}
_1 \varepsilon^0_a = \  \partial_a \ _0 \varepsilon^0, \quad \ _1
\varepsilon^0_3 = \ _0 \varepsilon^1, \quad   \ _1 \varepsilon^0 =
0,
\end{equation}
\begin{equation}\label{C7}
_1 \varepsilon^1_a  = \partial_a \ _0 \varepsilon^1, \quad \ _1
\varepsilon^1_3 = \ _0 \varepsilon^2,
\end{equation}
\begin{equation}\label{C8}
_0 \varepsilon^1_a = \  - \partial_a \ _0 \varepsilon^0_3, \quad \
_0 \varepsilon^1_3 = (1 - f')/2,
\end{equation}
\begin{equation}\label{C9}
_1 g'^0_{3a}=  \partial_a \ _1 \varepsilon^0_3 + \ _1
\varepsilon^1_a, \ _1 g'^0_{33}= 2 \ _1 \varepsilon^1_3, \ \ _1
g'^0_{aa}= 2 \partial_a \ _1 \varepsilon^0_a,
\end{equation}
where we have used the notation $\varepsilon^i \equiv
\varepsilon_i$.

A particular solution of this system is
\begin{equation}\label{C10}
_0 \varepsilon^1 = \ _1 \varepsilon^0 = 0, \ \ _1 \varepsilon^0_i
= \ _1 \varepsilon^1_i = 0, \ _0 \varepsilon^1_a =  -
\partial_a \ _0 \varepsilon^0_3,
\end{equation}
\begin{equation}\label{C11}
_0 \varepsilon^1_3 = (1 -f')/2, \quad \partial^2_{aa} \ _0
\varepsilon^0 = 0.
\end{equation}
On the other hand, we similarly obtain:
\begin{equation}\label{C12}
_0g'^1_{aa}= 2 \partial_a \ _0 \varepsilon^1_a
\end{equation}
which, according to the corresponding equation in (\ref{C10}),
becomes
\begin{equation}\label{C13}
_0g'^1_{aa}= - \partial^2_{aa} \ _0 \varepsilon^0_3.
\end{equation}
Thus, since $\ _0\varepsilon^0_3$ is small, but otherwise
arbitrary, we always can choose $\ _0\varepsilon^0_3$ so as to
have
\begin{equation}\label{C14}
\int \hspace{-2mm}\ _0 g'^1_{aa} \ dx^1 d x^2 \neq 0,
\end{equation}
that is, so as to have $P'^0 \neq 0$. Then, as we have announced,
we cannot preserve the vanishing of $P'^\alpha$ and $J'^{\alpha
\beta}$ when making a general coordinate transformation from an
intrinsic coordinate system to another intrinsic one.

\end{document}